\documentclass[a4paper]{aa}
\usepackage{psfig}
\usepackage{epsfig}
\usepackage{graphicx}

\def \inte {INTEGRAL}

\def \asca {AX~J161929--4945}
\def \src {IGR~J16195--4945}

\def \hcm {\hbox {\ifmmode $ atom cm$^{-2}\else atom cm$^{-2}$\fi}}

\def\approxgt{\mathrel{\hbox{\rlap{\lower.55ex \hbox {$\sim$}}
        \kern-.3em \raise.4ex \hbox{$>$}}}}
\def\approxlt{\mathrel{\hbox{\rlap{\lower.55ex \hbox {$\sim$}}
        \kern-.3em \raise.4ex \hbox{$<$}}}}

\begin{document}


\title{The soft X-ray counterpart of the newly discovered
INTEGRAL source  \src}

\author{L. Sidoli\inst{1}
        \and S. Vercellone\inst{1}
        \and S. Mereghetti\inst{1}
        \and M. Tavani\inst{2,3,4}}

\offprints{L.Sidoli (sidoli@mi.iasf.cnr.it)}

\institute{
        Istituto di Astrofisica Spaziale e Fisica Cosmica --
        Sezione di Milano ``G. Occhialini" -- IASF/CNR \\
         via Bassini 15, I-20133 Milano, Italy
\and
CIFS,Viale Settimio Severo 63, Torino, I-10133, Italy 
\and
        Istituto di Astrofisica Spaziale e Fisica Cosmica --
        Sezione di Roma -- IASF/CNR \\
         via Via del Fosso del Cavaliere 100, Roma, I-00133, Italy 
\and
Dipartimento di Fisica, Universit\`a di Tor Vergata, Roma, I-00133, Italy 
}

\date{Received 14 October 2004; Accepted: 22 November 2004 }

\authorrunning{L. Sidoli et al.}

\titlerunning{{The soft X--ray counterpart of the new $\Gamma$-ray source \src}}

\abstract{The \inte\ satellite, during its regular scanning
observations of the Galactic plane, has discovered several new
X-ray sources emitting above 20 keV. The nature of the great
majority of them is still unknown. Here we report on the likely
low energy counterpart, observed with ASCA in 1994 and 1997, of
one of these sources,  \src. The ASCA source is faint 
(F$_{\rm 2-10 keV}$$\sim$10$^{-11}$~erg~cm$^{-2}$~s$^{-1}$), highly absorbed
(N$_{\rm H}$$\sim$10$^{23}$~cm$^{-2}$) and has a rather hard
spectrum (photon index$\sim$0.6). These spectral properties are
suggestive of a neutron star in a High Mass X--ray Binary. Our
analysis of all the public INTEGRAL data of \src\ shows that this
source is variable and was in a high state with a 20-40 keV flux
of  $\sim$17 mCrab in two occasions in March 2003.
}

\maketitle

\section{Introduction}
\label{sect:intro}

Several  hard X--ray ($>$15 keV) sources have been discovered with
the INTEGRAL satellite (Winkler et al. 2003) during observations
of the Galactic plane performed in the last two years (e.g.,
Courvoisier et al. 2003, Walter et al. 2003; Rodriguez et al.
2003; Revnivtsev et al. 2003, see also Revnivtsev et al
2004, Molkov et al. 2004, Tomsick et al. 2004, 
Bird et al. 2004 and references therein for an updated
list). Observations at lower  X--ray energies are available for  a
few of these sources  and show that they are heavily absorbed
($>$10$^{23}$~cm$^{-2}$). The column densities are in some cases
larger than the values expected from the interstellar matter along
the line of sight, implying that the X--ray sources are embedded
in a local absorbing gas (Walter et al. 2003, Patel et al. 2004).

The high absorption, together with the transient or variable
nature of most of them, hampered their detection in previous
X--ray surveys carried out at lower energy. However, searches in
archival data
show that some of the INTEGRAL  sources are not new and do have
faint X--ray counterparts below 10 keV
(e.g. Rodriguez et al. 2003). The 2--10 keV X--ray spectra are
power laws with photon index $\sim$0.5--1, often with  strong iron
emission lines, as in the case of IGR~J16318--4848 (Matt \& Guainazzi, 2003).
The emerging
picture of this class of sources is that they are mostly High Mass
X--ray Binaries (HMXRBs). This is indicated, in a few cases, by
the identification of their optical/IR counterparts 
(e.g., Filliatre \& Chaty 2004, Negueruela \& Reig 2004) and/or by the
discovery of pulsations 
(e.g., IGR~J18027--2016/SAX~J1802.7--2017, 
Augello et al. 2003; 
IGR~J18410--0535/AXJ1841.0--0536, Halpern \& Gotthelf 2004).

However, many  new INTEGRAL sources, especially among the faintest
ones (a few milliCrabs) still lack  an identification and their
nature is  unknown. Thus, the study of their soft X--ray
counterparts is an important step for  unveiling their nature.

\src\ is one of the faint sources discovered with \inte\ during
observations carried out between February~27 and October~19, 2003
(Walter et al. 2004).
These authors reported it in a list of new sources, as a
10$\sigma$ detection in the 20--40~keV range, but
without any information on its flux level. In the catalogue of
Bird et al. (2004) \src\ is reported with a mean counting rate in
the IBIS instrument of 0.29$\pm{0.03}$~counts~s$^{-1}$  (20--40
keV) and an upper limit of 0.29 counts s$^{-1}$ (40--100~keV).
These values correspond approximately to 3 and 4~mCrabs in the
respective energy ranges.

We searched for possible lower energy counterparts of \src,
finding a faint X--ray source detected with the ASCA satellite in
the 2--10 keV band. Here we report the analysis of all the ASCA
observations and of the INTEGRAL public data of this source.

\section{ASCA Observations and Analysis}
\label{sect:asca_obs}

The sky coordinates of \src\ are 
R.A.= 16$^h$19.4$^m$, Dec.=-49$^\circ$ 43$'$ 08$''$ (J2000) 
with an uncertainty of 3$'$ (Bird et al. 2004).
Our search
in public archives of previous X--ray missions resulted in only
one object consistent with this position: 
the source \asca,
observed in 1994 and 1997 during the ASCA survey of the Galactic
Ridge (Sugizaki et al 2001; R.A. (J2000)=16$^h$19$^m$
29$^s$, Dec.=$-49^\circ45.5$$'$, 1$'$ error radius). 
Sugizaki et al (2001) report that
\asca\ is a  hard (or severely absorbed) source, not detected
below 2~keV.

No ROSAT  counterparts can be found in the RASS catalog,
consistent with a  very high absorption.

The region of sky containing \asca\ was observed three times
during 1994-1997 (see Table~1). The ASCA satellite
(Tanaka, Inoue \& Holt 1994) provides simultaneous data in four
co-aligned telescopes, equipped with two solid state detectors
(SIS) and two gas scintillation proportional counters (GIS).
However, since \asca\ was observed at a relatively high off-axis
angle, only the GIS instruments, which has a larger field of view
(44$'$ diameter), provide useful data.

We  reduced and analyzed the GIS data of the three observations,
obtaining for \asca\ the background subtracted count rates
reported in Table~1. These values, which include
the vignetting correction to account for the instrumental response
at different off-axis angles, give  evidence for long-term
temporal variability between 1994 and 1997, the source being
$\sim$2.5 times brighter in 1997. 
In Fig.~1 we
show the  light curve of the 1997 observation, where a
variability on shorter timescales of hours is clearly evident.

Only the 1997 observation provides enough statistics for a
meaningful spectral analysis. The source spectrum in the 1-10 keV
range is well fit by an absorbed power law model with photon index
$\Gamma$=0.6$^{+0.8} _{-0.5}$ and a relatively high value of N$_{\rm
H}$=12 ($^{+8} _{-4}$) $\times10^{22}$~cm$^{-2}$ (reduced
$\chi^2=1.12$, with 28 dof).
The observed flux is  $1.6\times
10^{-11}$~erg~cm$^{-2}$~s$^{-1}$, which becomes 
$2.4\times10^{-11}$~erg~cm$^{-2}$~s$^{-1}$ (2--10~keV) after correcting for
the absorption. 
There is no evidence for K$_{\alpha}$ iron lines, with
a 90\% level upper limit on the equivalent width of EW$<$270~eV
for a narrow line at 6.7~keV, and EW$<$470~eV for a narrow line at 6.4~keV.
We tried to fit the spectrum with other
simple models (blackbody, thermal bremsstrahlung, cut-off
powerlaw) but they were not adequate: the blackbody, although
formally acceptable (reduced $\chi^2=1.22$, with 28 dof), resulted
in a very high temperature, kT, of $\sim$3~keV, leaving strucured 
residuals. Both the  bremsstrahlung and the cut-off
power law resulted in an unconstrained cut-off (or temperature) at
high energy.

A search for periodic pulsations in the 0.1-1000 s range using the
data of the 1997 observation gave negative results.

\begin{figure}[!ht]
\centerline{\psfig{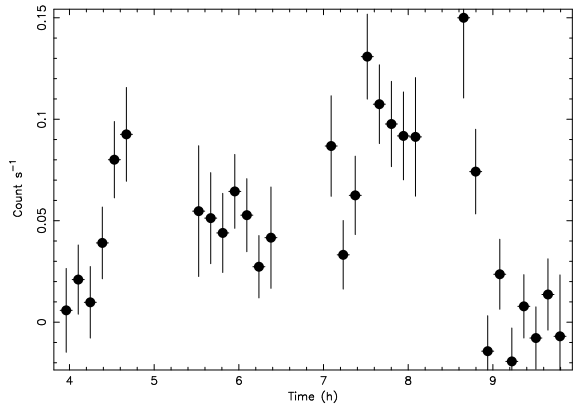}}
\vskip 0.0truecm \caption{{Background subtracted light curve
obtained from the sum of the two GIS instruments during the 1997
observation. Time is in hours from the beginning of the
observation (3 Sep 1997, 03:01:20). Bin size is 512~s. The source is variable
and goes below the threshold of detectability at the beginning and at the end
of the observation.
}}
 \label{fig:asca_lc}
\end{figure}


\begin{table}[htbp]
\label{tab:ascaobs}
\begin{center}
  \caption{Summary of the ASCA observations of \asca. The count rates are
corrected for the vignetting.
}
    \begin{tabular}[c]{cccll}
\hline
 Start date & GIS Exp. &  Count rate & Off-axis \\
         (UT)         & time (ks)  & (10$^{-2}$ cts s$^{-1}$) & angle \\
\hline
 01 Sep 1994  &  4.0  & 6.3 $\pm{0.7}$ &  14$'$   \\
 02 Sep 1994  &  7.2  & 7.0 $\pm{1.1}$ &  21$'$   \\
 03 Sep 1997  & 12.6  & 17.3 $\pm{0.8}$ &  21$'$   \\
\end{tabular}
\end{center}
\end{table}


\begin{figure}[!ht]
\centerline{\psfig{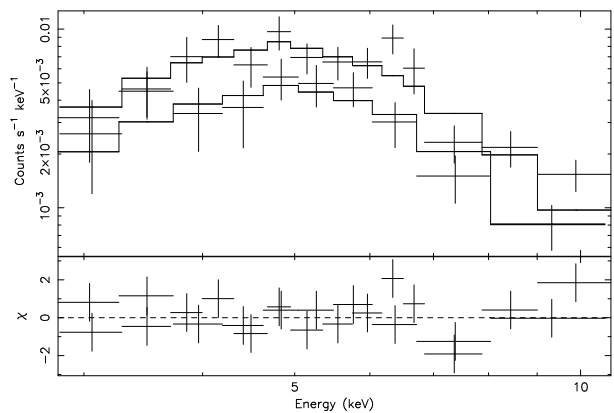}}
\vskip 0.0truecm \caption{{Best fit power law spectrum of \asca\
obtained with the GIS2 and GIS3  instruments. The residuals in
units of standard deviations are shown in the lower panel. }}
\label{fig:gis_spec}
\end{figure}


\begin{figure}[!ht]
\centerline{\psfig{figure=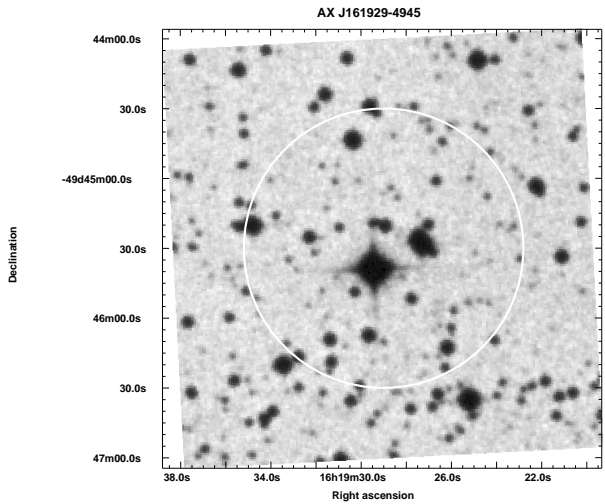,height=8.0cm,angle=0}} \vskip
0.0truecm \caption{{R-band image of the error region of \asca\
obtained from the ESO-R survey scanned on the Precision Measuring Machine
at the US Naval Observatory, Flagstaff Station.
The circle has a radius of 1$'$. North is to
the top, East to the left. The brightest star in the error circle
is HD 146628 (V=10.08).}} 
\end{figure}

\section{Discussion}
\label{sect:inte_obs}

It is very likely that \asca\ is the low energy counterpart of
\src. Its spectral properties are similar to those of other
highly absorbed sources which have been associated to the newly
discovered INTEGRAL sources. Furthermore, based on the galactic
LogN-LogS in the 2--20~keV energy band (Sugizaki et al 2001),
the chance probability of finding a
source brighter than $\sim$10$^{-11}$~erg~cm$^{-2}$~s$^{-1}$ in the 3$'$ 
radius error box is $\sim 2.4\times 10^{-3}$.

Extrapolating the 2--10 keV best fit power law  to higher energy,
we obtain a 20--40~keV flux of 
$1.2\times10^{-10}$~erg~cm$^{-2}$~s$^{-1}$. 
This is higher than the average
flux of $\sim$3 mCrab reported  for \src\ (Bird et al. 2004).
However the latter is an average value derived from all the
INTEGRAL data taken over the period between 2003 February 28 
and October 10.

To investigate in more detail the variability properties of the
source, we analyzed all the INTEGRAL observations of this region of
the sky which are currently public.
We restricted our analysis to the data of
the IBIS instrument (Ubertini et al. 2003) for observations pointed at
less than 5$^{\circ}$ from the source position.
This selection resulted in 56  observations obtained 
in March--April 2003, 
with an approximate total net exposure time of $\sim$100~ksec. 
The \inte\
data have been reduced with the standard procedure using version 4
of the ISDC Offline Scientific Analysis software (OSA, Courvoisier
et al. 2003b).

Only in 2  observations of $\sim$1.7~ks each (see Table~2), 
the IBIS/ISGRI detection significance was
above 3~$\sigma$ in the 20--40~keV energy band, at a flux level of
$\sim$17 mCrab.
\src\ was never detected above 40~keV. These flux
values are more in line with those measured by ASCA (see Fig.~4
for the broad-band non-simultaneous spectrum). The source is
clearly variable at soft and hard X-ray energies, and only during
the brightest flares it was above the INTEGRAL sensitivity.

\begin{table}[htbp]
\label{tab:inteobs}
\begin{center}
  \caption{Summary of the two \inte\ observations where a significant 
detection for \src\ has been obtained. IBIS/ISGRI count rates are
in the 20--40~keV energy band }
    \begin{tabular}[l]{lll}
\hline
\inte\ & Start Time      &  ISGRI rate \\
Obs.   &   (UT)          & (s$^{-1}$) \\
\hline
004700970010 & 05 Mar 2003  08:15:35  & 1.78 $\pm{0.45}$   \\
005000770010 & 14 Mar 2003  00:30:54  & 1.60 $\pm{0.43}$  \\
\end{tabular}
\end{center}
\end{table}


\begin{figure}[!ht]
\centerline{\psfig{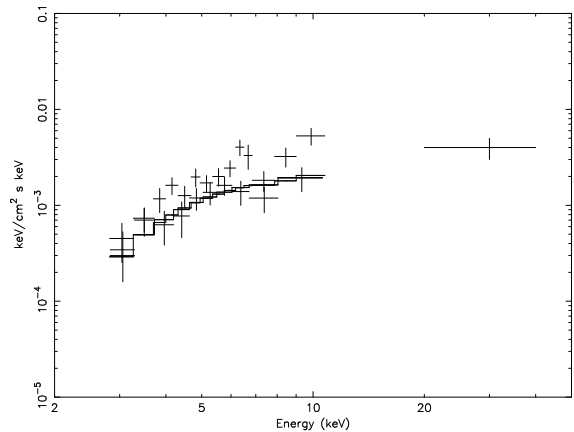}}
\vskip 0.0truecm \caption{{Broad band Ef(E) 
spectrum obtained with the ASCA best-fit together
with the IBIS/ISGRI flux  measured for \src\ in the energy 
band 20--40~keV (see Table~2).
}}
\label{fig:gis_spec}
\end{figure}

An R-band image of the error region of \asca\ is shown
in Fig.~3. The brightest star in the error circle is 
HD~146628 (B=10.52, V=10.08).
Different spectral types and
possible distances for this star are reported in literature: 
it is classified as a B-type supergiant 
(B1/B2Ia sp.type, HD~catalogue, Cannon \& Pickering 1918-1924),
located at about 7~kpc (distance calculated 
assuming canonical relations reported
in Allen et al., 2000), or as a O-type main sequence star at $\sim$2~kpc 
(O9V sp.type, M$_{V}$=$-4.54$, A$_{V}$=2.78, Kilkenny 1993).
A distance of few kpc implies a 2--10 keV luminosity 
of $\sim$10$^{35}$~erg~s$^{-1}$, still compatible with being 
a galactic X--ray binary.
A bright 2MASS infrared source is positionally 
coincident with HD~146628 
(J=$8.575\pm{0.026}$, H=$8.412\pm{0.042}$ and K=$8.377\pm{0.031}$).

The total Galactic absorption in this direction is 
N$_{\rm H}$$\sim$2$\times$10$^{22}$~cm$^{-2}$ (Dickey \& Lockman, 1990).
As expected, the
optical absorption derived for HD~146628 (A$_{V}$=2.78, Kilkenny 1993)
corresponds to a lower column density value 
(N$_{\rm H}$$\sim$5$\times$10$^{21}$~cm$^{-2}$; 
Predehl \& Schmitt, 1995).
The fact that this value is smaller than that measured for the ASCA source
does not rule out the possible  association of HD~146628
with the X--ray source, if we admit that the latter, 
maybe enshrouded by a dense local envelope,
is intrinsically much more absorbed than its companion.
Indeed, this is also suggested by the fact that the absorption of \asca\ is
even larger than the total Galactic absorption in this direction.

The fact that \asca\ has been always detected with ASCA GIS both in
1994 and in 1997 suggests a persistent, although quite variable, nature
for the high energy emission. All the source properties, the hard
spectrum, the strong absorption, the temporal variability are
strongly suggestive of a neutron star in a High Mass X--ray
Binary.

Follow-up observations at  X--rays are needed to improve the
source position in order to establish if the star is really
physically related to the high energy source.

\begin{acknowledgements}
This research has made use of data obtained through the public
INTEGRAL Data Archive provided by the INTEGRAL Science Data
Center, and of $ASCA$ data obtained through the High Energy
Astrophysics Science Archive Research Center Online Service,
provided by the NASA/Goddard Space Flight Center.
This research has made use of the USNOFS Image and Catalogue Archive
operated by the United States Naval Observatory, Flagstaff Station.
This publication makes use of data products from the Two Micron All Sky Survey, which is a joint project of the University of Massachusetts and the Infrared Processing and Analysis Center/California Institute of Technology, funded by the National Aeronautics and Space Administration and the National Science Foundation.
We would like to
thanks Ada Paizis, Diego Gotz and Masha Chernyakova 
for their help with the \inte\ reduction and data
analysis. This work has been supported by the Italian Space
Agency.
\end{acknowledgements}


\begin{thebibliography}{}


\bibitem[]{}
Allen's Astrophysical Quantities, 4th Edition, AIP Press, 2000

\bibitem[]{}
Augello, G., Iaria, R., Robba, N.R., et al. 2003,  ApJ, 596, L63 


\bibitem[]{}
Bird, A.J., Barlow, E.J., Bassani, L., et al. 2004, ApJ, 607, L33

\bibitem[]{}
Cannon, A.J.,  Pickering, E.C. 1918-1924, 
The Henry Draper Catalogue, Ann. Astron. Obs. Harvard College, 91-99


\bibitem[]{}
Courvoisier, T.J.-L., et al. 2003, IAU Circ., 8063

\bibitem[]{}
Courvoisier, T.J.-L., et al. 2003b, A\&A 411, L53

\bibitem[]{}
Dickey, J.M., Lockman, F.J., 1990, ARA\&A, 28, 215

\bibitem[]{}
Filliatre, P., Chaty, S., 2004, ApJ in press, astro-ph/0408407 


\bibitem[]{}
Halpern J.P., Gotthelf, E.V., 2004, ATel, 341

\bibitem[]{}
Johnson, H.L., 1966, ARA\&A, 4, 193

\bibitem[]{}
Kilkenny, D., 1993, SAAOC., 15, 53

\bibitem[]{}
Matt, G. \& Guainazzi, M., 2003, A\&A, 341, L13

\bibitem[]{}
Molkov, S.V., Cherepashchuk, A.M., Lutovinov, A.A., 2004, Astronomy Letters, 30, 534


\bibitem[]{}
Negueruela, I., Reig, P., 2004, ATel  285


\bibitem[]{}
Patel, S.K., Kouveliotou, C., Tennant, A., 2004, ApJ, 602, L45

\bibitem[]{}
Revnivtsev, M.G., et al. 2003, IAU Circ. 8097

\bibitem[]{}
Revnivtsev, M.G., et al. 2004, Sunyaev, R. A., Varshalovich, D. A., et al., 2004, Astronomy Letters, 30, 382



\bibitem[]{}
Rodriguez, J., Tomsick, J.A., Foschini, L., et al. 2003, A\&A, 407, L41

\bibitem[]{}
Smith, D.M., 2004, ATel, 338


\bibitem[]{}
Sugizaki, M., Mitsuda, K., Kaneda, H., et al 2001, ApJSS, 134, 77

\bibitem[]{}
Tanaka, Y., Inoue, H., Holt, S.S.,1994, PASJ 46, L37

\bibitem[]{}
Tomsick, J.A., Lingenfelter, R., Corbel, C., et al. 2004, Proc. V INTEGRAL Workshop, in press (astro-ph/0404420)

\bibitem[]{}
Ubertini, P., Lebrun, F., Di Cocco, G., et al. 2003, A\&A, 411, L131


\bibitem[]{}
Walter, R., Rodriguez, J., Foschini, L., et al., 2003, A\&A, 411, L427



\bibitem[]{}
Walter, R., et al., 2004, ATel 229



\bibitem[]{}
Winkler, C., Courvoisier T.J.-L., Di Cocco, G., et al., 2003, A\&A 411, L1



\end{thebibliography}
\end{document}